# Ratchet based ion pumps for selective ion separations

Alon Herman[1], Joel W. Ager[2,3], Shane Ardo[4], and Gideon Segev[1*]

[1]School of Electrical Engineering, Tel Aviv University, 6997801, Israel

[2]Department of Materials Science and Engineering, University of California at Berkeley, Berkeley, CA, 94720, USA

[3]Materials Sciences Division, Lawrence Berkeley National Laboratory, Berkeley, CA, 94720, USA

[4]Department of Chemistry, Department of Chemical & Biomolecular Engineering, Department of Materials Science & Engineering, University of California Irvine, Irvine, CA USA

[*]email: gideons1@tauex.tau.ac.il

**ABSTRACT**

The development of a selective, membrane-based ion separation technology may prove useful in a wide range of applications such as water treatment, battery recycling, ion specific chemical sensors, extraction of valuable metals from sea water, and bio-medical devices. In this work we show that a flashing ratchet mechanism can be used for high precision ion separation. The suggested ratchet-based ion pumps utilize a unique feature of ratchets, the frequency dependent current reversals, to drive ions with the same charge but different diffusion coefficients in opposite directions. We show that for the prevalent ions in water, ions with a relative diffusion coefficient difference as small as 1% can be separated by driving them in opposite directions with a velocity difference as high as 1.2 mm/s. Since the pumping properties of the ratchet are determined by an electric input signal, the proposed ion pumps can pave the way for simple large-scale, fit-to-purpose selective ion separation systems.

INTRODUCTION

In recent years there has been a growing demand for membranes capable of extracting a single solute form water.[1–3] The introduction of an ion selective separation technology can drive a dramatic progress in a range of applications such as water treatment, resource extraction from sea water, bio memetic systems, and chemical sensors. For example, according to the world health organization, the acceptable concentration of lead in drinking water is in the order of several particles per billion[4] and the concentration of NaCl in drinking water is slightly lower than one particle per thousand. Thus, a water decontamination system based on driving ions through a medium with a solute dependent permeability, should be about six orders of magnitudes more permeable to lead than towards sodium ions.

One of the main approaches pursued for solute-solute selective separation is filtration with sub-nanometer pore membranes. In these membranes, the pore diameter is engineered to be between the diameter of the hydrated target ion and its diameter when it is partially dehydrated. Ion transport through the membrane is achieved by applying energy to partially (or fully) dehydrate ions and transport them through the membrane. Selectivity originates from the difference in the hydration energy between ions and the difference in their transport properties within the pores.[1,2] Although achieving some success, highly selective separation of same-charge ions remains a fundamental challenge.[1,2] Furthermore, since these membranes rely on functionalized pores with a precise diameter below 1 nm, upscaling these membranes is a great challenge. Moreover, as the membrane selectivity is derived by the pore geometry and chemistry, unique membranes must be designed for every target ion.[2] Another approach for monovalent ion separation is based on the diffusion coefficient dependent response of ions to a pressure induced steaming potential.[1,5–9] However, only a few studies have demonstrated this mechanism and so far their reported selectivity was limited. A separation system that can transport ions with the same charge in opposite directions might be able to bypass the limitations faced by current approaches.

Here we propose to use a ratchet mechanism to separate ions with the same charge number according to their diffusion coefficients. Since the selectivity is controlled by the ratcheting process, there is no need for energy intensive ion dehydration and complex sub-nanometer porous structures. Ratchet based separation harnesses a unique characteristic of ratchets to drive particles of different diffusion coefficients in opposite directions, even when having the same charge. This ability paves the way towards removal of solutes found in trace levels in mixtures, which is a fundamental challenge in ions separations. Furthermore, since the separation properties are controlled by an electrical input signal, ratchet-based separation may pave the way to simple fit-to-purpose and real-time applications.

Electronic flashing ratchets are non-equilibrium devices that utilize modulation of a spatially asymmetric electric field to drive a steady state particle flux.[10–12] Similar to a MOSFET transistor, most of the demonstrated flashing ratchets are comprised of gating electrodes that generate a modulated electric field, and source and drain electrodes between which the ratchet output current flows. The symmetry breaking required to produce a directed current is achieved by structuring the gating electrodes,[13–17] introducing a nonlinear, diode like conductance between the source and drain,[18,19] or by applying different signals to different sets of gating electrodes.[20–26] Figure 1a shows a schematic illustration of the operating principles of a flashing ratchet driving positive charge.[13,27] The empty circles represent the ions position at the beginning of every step, and the filled circles represent their position at the end of every step. The potential distribution through the device ($V$, solid blue line) is switched between two saw-tooth potential distributions $V^+$ and $V^-$ where $V^- = \alpha V^+$, and the potential asymmetry factor $\alpha$ is negative. Initially the particles rest at the potential minima, for example at $x_1$. When the potential is switched to $V^-$ at $t=t_0$, the particles drift in both directions. The length of the time interval between $t_0$ and $t_1$ is tuned to allow some drifting particles to reach the potential minimum on their right, $x_1'$, but not the potential minimum to their left, $x_0'$. Hence, when the potential is switched to $V^+$ at $t=t_1$, some of the particles at $x_1'$ drift further to the right towards $x_2$, but all the other particles return to their initial position at $x_1$ resulting in a net current to the right. The asymmetric potential distribution allows some particles to reach the potential minima in one direction before particles drifting in the opposite direction, thus making it essential for a non-zero net current. Figure 1b illustrates the operation of the same ratchet driving negatively charged particles. Here particles drift towards the potential peaks and as a result, the same potential distribution and input signal generate a net particle flux in the opposite direction.

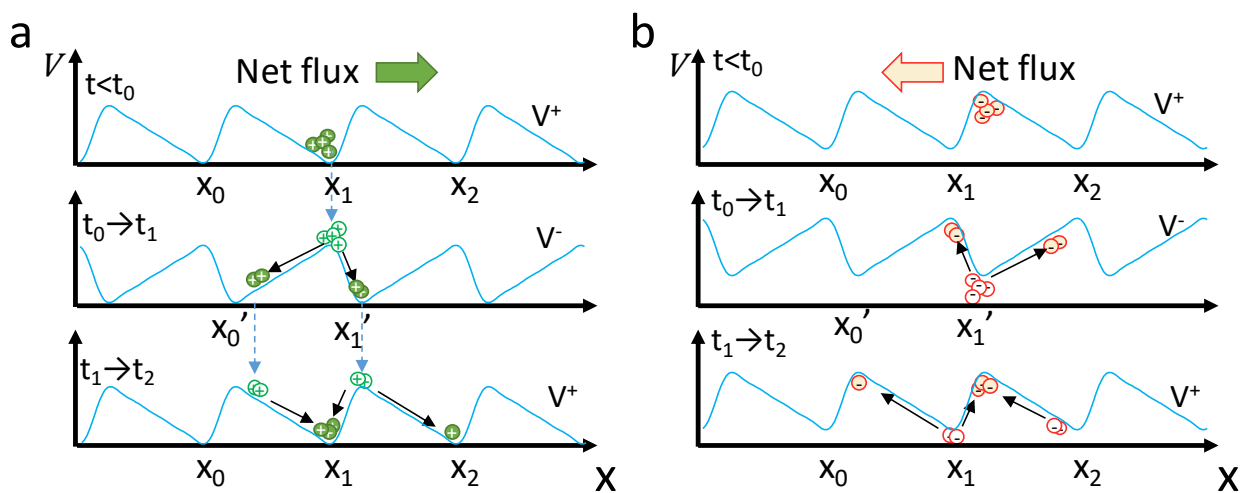

*Figure 1: the operating principles of a flashing ratchet driving positively and negatively charged particles (a and b respectively). The solid blue line illustrates the potential distribution during sequential steps, empty circles mark the position of particles at the beginning of every step and filled circles mark their position at the end of every step.*

While in most prior studies flashing rachets were used to drive electrons or holes in semiconductors,[13–26] the application of flashing ratchets for ion pumping was only suggested recently. Membranes with oscillating pores for resonant osmosis were theoretically analyzed in terms of ratchet transport[28,29] and nanofluidic charge coupled devices, which can be viewed as reversible ratchets,[30] were studied for DNA separations.[31] In prior work, we have utilized these ratcheting principles to demonstrate experimentally a first-of-its-kind ratchet-based ion pump.[32] In all analyses of ratchet based ion pumps (RBIPs) for ion separation,[28,29,31] selectivity is achieved by transporting particles with different velocities according to their diffusion coefficients. However, driving charged particles in the same direction but different velocities meet the same selectivity limitations discussed above when attempting to remove trace concentrations of the target ion from a mixture.

An important hallmark of ratchets is the ability to invert the direction of particle flow with a change in the input signal frequency.[13,33] The stopping frequency, which is the input signal frequency at which the particle flux changes its direction, is determined by the potential distribution and particles transport properties.[34–36] As a result, for a given ratchet, there can be a frequency at which particles with the same charge but different diffusion coefficients are transported in opposite directions. This effect was used to sort gold nanoparticles of different sizes and shapes[35–37] and was studied theoretically in ratchets based on pulsating the spacing between the walls of asymmetric periodic channels or their relative position.[38–42] This concept was never applied to ion separations. Here we show that by utilizing velocity reversal in ratchet systems, ions with the same charge can be driven in opposite directions according to their diffusion coefficients. This enables extraction of ions with extremely low relative concentrations if their diffusion coefficient is even slightly different from the main ions in the solution. Since the direction of ion transport is determined by the input signal frequency, the sorting properties of the ions can be tuned in real time providing a simple fit-to-purpose solution for a variety of ion separations applications.

**Analysis**

In the absence of bulk chemical reactions, the transport of noninteracting ions in a solution is determined by the continuity equation:

$$\frac{\partial C(x,t)}{dt} = -\frac{\partial}{dx} N(x,t) \qquad (1)$$

where $C(x,t)$ is the ion concentration, and $N(x,t)$ is the ion flux. When convection is negligible, the ion flux is driven by drift and diffusion processes:

$$N(x,t) = -D\left[z\beta C(x,t)\frac{\partial}{\partial x} U(x,t) + \frac{\partial}{\partial x} C(x,t)\right] \qquad (2)$$

Here $D$ is the ion diffusion coefficient, $z$ is the ion charge number, $\beta = e/k_B T_r$ is the inverse thermal voltage, $k_B$ is the Boltzmann constant, $T_r$ is the temperature, and $e$ is the elementary charge. The potential distribution in the system $U(x,t)$ is described as the product of spatial and temporal components:

$$U(x,t) = V(x)g(t) \qquad (3)$$

We address in this paper a simple ratchet system that has an infinitely periodic sawtooth distribution in space, with a spatial period $L$, and is driven by a periodic square wave signal at frequency $f$. The spatial and temporal contributions are shown in Figure 2, and are mathematically described by:

$$V(x) = \begin{cases} V_{max}\dfrac{x/L}{x_c}, & 0 < \dfrac{x}{L} < x_c \\ V_{max}\dfrac{(1-x/L)}{(1-x_c)}, & x_c < \dfrac{x}{L} < 1 \end{cases} \quad ; \quad g(t) = \begin{cases} 1, & 0 < t < \delta \cdot T \\ \alpha, & \delta \cdot T < t < T \end{cases} \qquad (4)$$

Where $V_{max}$ is the maximum potential amplitude, and $x_c$ is the relative length of the first linear section of the sawtooth potential. The temporal modulation $g(t)$ is described by a duty-cycle $\delta = t^+/T$, which is the ratio between the time-duration of the 1st step, where the potential is at its maximum value, to the total period $T(=1/f)$. Each time-period is completed with a 2nd step, in which the sawtooth distribution is multiplied by a potential symmetry factor in the range $-1 \leq \alpha \leq 0$.

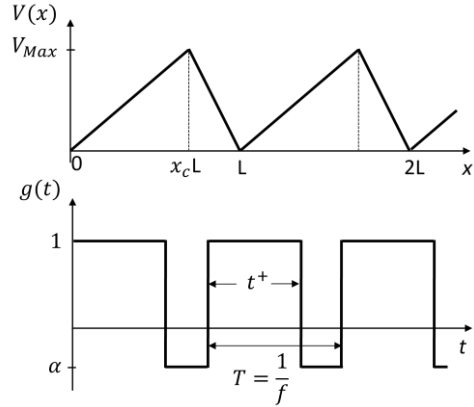

Figure 2. The spatial and temporal components of the potential distribution

The continuity equation was solved numerically using COMSOL® Multiphysics v5.5. The model domain is $x \in [0, L]$, with a periodic boundary condition defined as $C(0,t) = C(L,t)$, and a normalization condition for the total concentration $\int_0^L c(x,t)dx/L = c_0$, where $c_0$ is a reference ion concentration. The time-dependent simulation runs over several consecutive temporal periods, until the net ion flux reaches its steady state value, and the ratchet operation is independent of its initial condition. We define the time $\tau = 0$ to be the time at which the system enters its steady state. Thus, the ratchet steady state operation is studied by analyzing the interval $0 < \tau < T$. Our analysis is presented in terms of ion velocity (and not flux), since it is independent of the reference concentration, and therefore is a more general characteristic of the ratchet. We define the instantaneous mean ion velocity, $\bar{v}(\tau)$, and the net ion velocity, $v_{net}$, to be the average velocity in space and time, respectively:

$$\bar{v}(\tau) = \frac{1}{c_0 L}\int_0^L N(x,\tau)\,dx \quad (5)$$

$$v_{net} = \frac{1}{T}\int_0^T \bar{v}(\tau)d\tau \quad (6)$$

The model predictions were verified by comparing them to prior analytical results under similar conditions.[13,27] More information on the model validation can be found in the SI.

## Results

### Velocity reversal

Ratchets can be useful for many applications, but of particular interest to us is to examine the range of parameters that drive ions with the same charge (polarity and valence), in opposite directions. The ion response to a ratchet drive was found by calculating the net ion velocity, $v_{net}$, according to eq. (5)-(6) for various input signals, $g(t)$, potential distributions, V(x), and ions with diffusion coefficients, D. Unless stated otherwise, the baseline ratchet parameters that are used in the analysis are: $L = 1$ µm, $x_c = 0.7$, $V_{max} = 2.5$ V, $\beta = 39.59$ V$^{-1}$, $\delta = 0.25$, $\alpha = -0.5$, and the ion charge number is $z = 1$. Figure 3 illustrates a typical velocity reversal example. The figure shows the net ion velocity as a function of the ratchet signal frequency for two ions with different diffusion coefficients. For input signal frequencies between 62–103 kHz, the low $D$ ions have a positive net velocity, and the high $D$ ions have a negative net velocity.

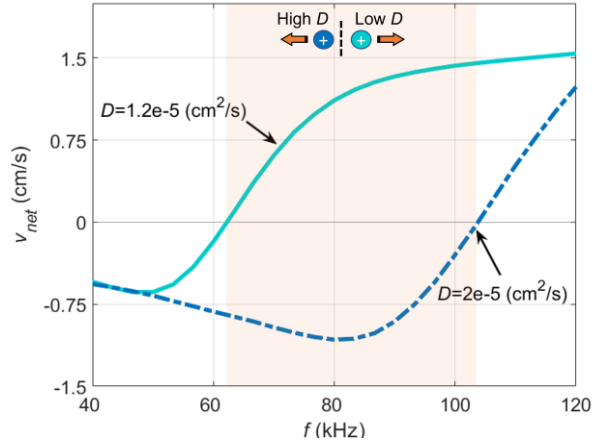

Figure 3. The net velocity of two positively charged monovalent ions ($z = 1$), as a function of the input signal frequency. In the shaded region the ions move in opposite directions. The ratchet parameters are: $L = 1$ µm, $x_c = 0.7$, $V_{max} = 2.5$ V, $\delta = 0.25$, $\alpha = -0.5$.

The ion distribution is shown in Figure 4 at a few key times along a cycle, and at a frequency of 83 kHz, corresponding to the largest difference in velocities in Figure 3. Figure 4a describes the evolution of the low $D$ ions and Figure 4b describes the evolution of the high $D$ ions. Starting from $\tau = 0$, we see that the entire ion population is crowded around the potential minima. After the 1$^{st}$ switch (from negative to positive potential) at $\tau = 0.02T$, the ion population is split into two groups drifting in opposite directions. By the time of the 2$^{nd}$ potential switch (from positive to negative potential) at $\tau = 0.25T$, nearly all the high $D$ ions (Figure 4b) have reached the new potential minima, while the majority of the low $D$ ions (Figure 4a) have not. After the 2$^{nd}$ switch, at $\tau = 0.26T$, the ions split again, but since the electric fields are lower ($\alpha = -0.5$), the groups drift apart more slowly. This allows more ions to diffuse over the potential peak and join the ions drifting in the negative x direction. When the groups are completely separated, at $\tau = 0.28T$, this process stops. The overall effect is a transfer of ions from the low field side of the sawtooth to the high field side. We call this the *injection* phenomenon. For the high $D$ ions (Figure 4b), the injection from right to left is the main reason for the net negative velocity: after the 1$^{st}$ switch 57% are traveling to the left, while after the 2$^{nd}$ switch only 45% return to the right. Hence, in every time-period, about 12% of the ions are moving one spatial period to the negative direction, resulting in a net velocity of $\bar{v} = -12\% \cdot L/T = -1$ cm/s. For the low $D$ ions (Figure 4a) however, there is a large group that does not reach the potential minimum before the 2$^{nd}$ switch, that is joined by a smaller group after the 2$^{nd}$ switch, to travel back right. Overall, 75% of the ions are returning to the right ($\tau = 0.28T$), compared to 60% that traveled left after the 1$^{st}$ switch ($\tau = 0.02T$), resulting in a net velocity of about $\bar{v} = 15\% \cdot L/T = 1.25$ cm/s. We note that injection also occurs with the low $D$ ions near the potential maxima after the 2$^{nd}$ switch ($\tau = 0.26T$), but since only a minority of the ions have reached there, this injection is much less significant. At $\tau = 0.5T$, the ions that travelled in the high field side of the sawtooth reach the potential minima; and by the end of the cycle, all the ions reach potential minima, and a new cycle starts.

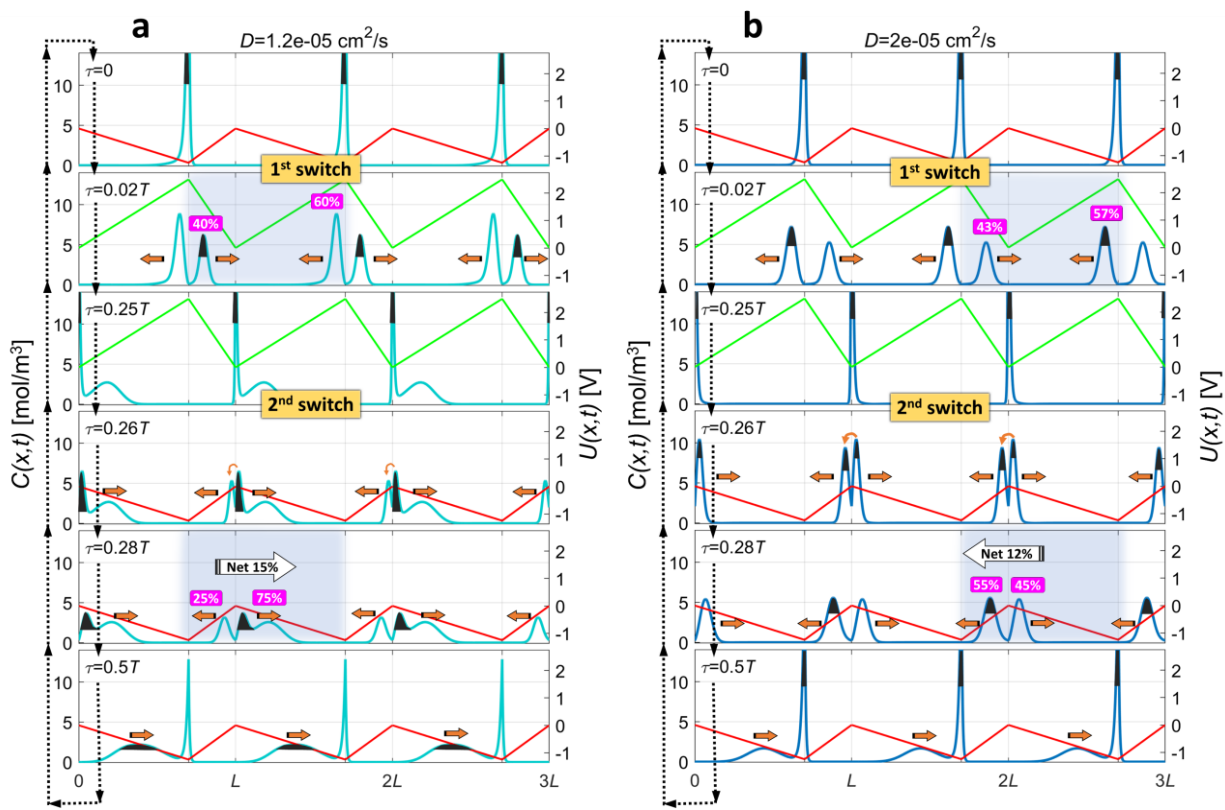

Figure 4. The concentration response $C(x,t)$ of two positively charged monovalent ions ($z = 1$) with diffusivities $D = 1.2 \cdot 10^{-5}$ $cm^2/s$ (a), and $D = 2 \cdot 10^{-5}$ $cm^2/s$ (b). The ratchet characteristics are presented in Figure 3, and the frequency is f = 83kHz. The right axis in both graphs shows also the sawtooth potential distribution $U(x,t)$.

Figure 5 shows a 2D map of the net velocity of monovalent cations as a function of diffusion coefficient and the input signal frequency. All other ratchet ratchet parameters are as in Figure 3 – Figure 4. It can be easily noticed that the stopping frequency, which is the frequency at which the net velocity is zero, increases with the diffusion coefficient. This implies that for a given frequency, ions with diffusion coefficients above a specific value have a positive velocity, and ions with diffusion coefficients below that value travel in the opposite direction. For example, at a frequency of 60 kHz, K$^+$ and Na$^+$ ($D = 1.96 \cdot 10^{-5}$ and $D = 1.33 \cdot 10^{-5}$ $cm^2/s$ respectively) travel backward with velocities of about –0.8 and –0.5 cm/s respectively, but Li$^+$ ($D = 1.03 \cdot 10^{-5}$ $cm^2/s$) travels forwards at a velocity of +0.6 cm/s. This demonstrates the ability of ratchets to separate ions by driving them in opposite directions according to their diffusion coefficients. Moreover, since the input signal determines which ions move forward and which move backward, the ionic selectivity of these ratchets can be tuned in real time.

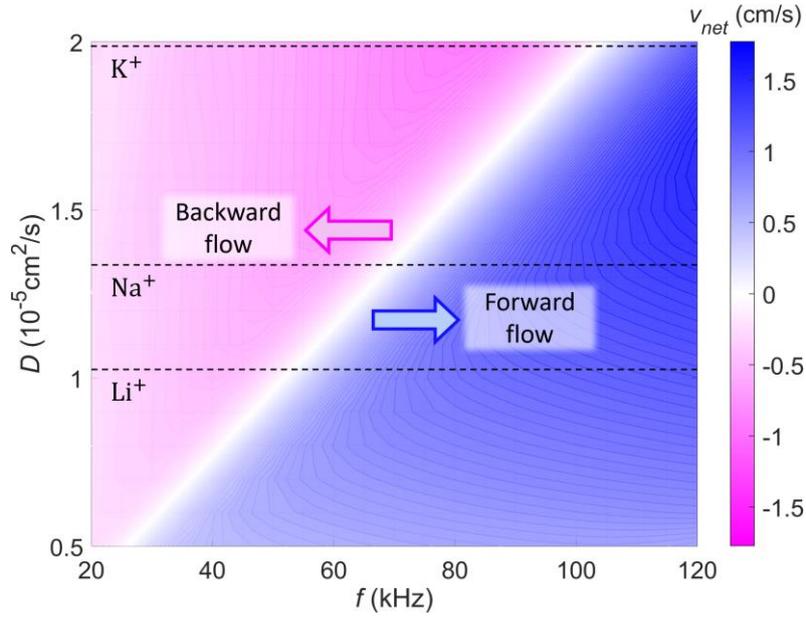

*Figure 5. Net velocity of monovalent cations as a function of the input signal frequency and ion diffusion coefficient. All potential and signal parameters are the same as in Figure 3 Figure 4.*

To analyze the ratchet behavior for a wide range of ratchet parameters, it is useful to normalize the net ion velocity and the signal's frequency by a characteristic ion velocity parameter $v_0 = zD\beta V_{max}/L$ and a characteristic ratchet frequency $f_0 = v_0/L$,[33] respectively. For convenience, a normalized frequency parameter $\Gamma = f/f_0$ was also defined. Figure 6a shows the normalized net ion velocity as a function of $\Gamma$ with different symmetry factors $\alpha$. Since the values are normalized, a single curve contains information about any diffusion coefficient or ratchet period. The inset shows the normalized net ion velocity for a wider $\Gamma$ range, illustrating a known ratchet property, that at very low and very high frequencies the output of a ratchet is zero.[33] For moderately low $\Gamma$ the behavior is similar to the operation shown in Figure 4b, where the negative net velocity is a result of the injection phenomenon described above. As $\alpha$ decrease in magnitude, the injection becomes more dominant, and therefore the net velocity becomes more negative. At the moderately high $\Gamma$ regime, the sign and magnitude of the velocity are determined by the time-averaged potential $\bar{U} = \delta V_{max} + (1-\delta)\alpha V_{max}$. For negative $\bar{U}$ ($\alpha < -0.33$) the net velocity is positive, and therefore a stopping frequency exists, while for positive $\bar{U}$ ($\alpha > -0.33$) the net velocity is negative, and there is no stopping frequency, and therefore no velocity reversal.

Figure 6b shows the normalized net ion velocity as a function of $\delta$ for moderately high $\Gamma$, with different symmetry factors $\alpha$. As expected, at $\delta = 0$ and $\delta = 1$, there is no net movement of ions since there are no temporal fluctuations. At $\alpha = -1$, the potentials at the positive and negative periods are equal in magnitude, therefore the velocity curve is antisymmetric with respect to $\delta = 0.5$. These 'sine' shapes are well known in ratchet systems,[13,27,33] and are a good validation of our model.

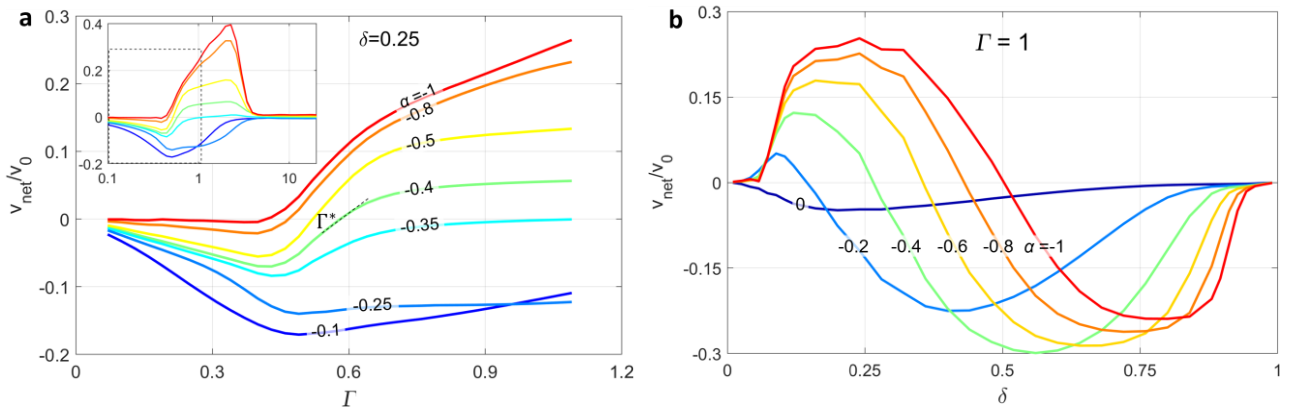

*Figure 6. Normalized net ion velocity characteristic curves for ion $z = 1$, $D = 1 \cdot 10^{-5}\ cm^2/s$, and ratchet parameters: $L = 1\ \mu m$, $x_c = 0.7$, for different symmetry factors. (a) $\delta = 0.25$, $V_{max} = 2.5\ V$, as function of the normalized frequency. (b) $\Gamma = 1$, $V_{max} = 1\ V$, as function of duty cycle.*

## The separation resolution

The selectivity of membranes for different same-charge ions is typically expressed in terms of the ratio of fluxes of the desired and undesired species crossing the membrane.[43,44] A highly selective membrane will yield a flux ratio on the order of 100:1 for the wanted species, and typically much lower than that.[44–48] In the case where same-charge ions are transported in opposite directions, none of the unwanted ions will cross the membrane, and this selectivity definition ceases to be relevant. Therefore, a new parameter $\mathcal{R}$ is introduced, called the *separation resolution*. It is defined as the ability to separate different same-charge ions according to the difference in their diffusion coefficients. According to this definition, a ratchet membrane can separate two types of ions with a diffusion coefficient difference of $\Delta D$, by driving one ion through the membrane at a velocity $v_{net} = \mathcal{R}\Delta D$, while the other ion is kept at its stopping frequency and is not affected by the ratchet. Since this definition describes a linear relation between the difference in diffusion coefficients and the ion net velocity, it becomes less accurate as $\Delta D$ increases.

The separation resolution for a certain set of ratchet parameters is calculated according to eq. (7) based on the normalized stopping frequency $\Gamma^*$, and the slope of the characteristic net velocity function $v_{net}/v_0 = \mathcal{H}(\Gamma)$, as shown for example in Figure 6a (the SI provides the detailed calculation of eq. (7)).

$$\mathcal{R}(\alpha, \delta, x_c, \beta V_{max}, L, z) = \left.\frac{\partial v_{net}}{\partial D}\right|_{\Gamma^*} = -\frac{z\beta V_{max}}{L} \cdot \Gamma^* \left.\frac{\partial \mathcal{H}}{\partial \Gamma}\right|_{\Gamma^*} \tag{7}$$

Figure 7 shows the separation resolution for different symmetry factors $\alpha$ and duty-cycles $\delta$. The light blue region corresponds to input signals that do not yield velocity reversal, showing more generally that for sawtooth sharpness $x_c > 0.5$, velocity reversal is possible only with $\bar{U} < 0$ (for $x_c < 0.5$ velocity reversal is possible only for $\bar{U} > 0$). High separation resolution is achieved using mid-range symmetry factors ($-0.8 < \alpha < -0.4$,) and mid-range duty-cycles ($0.2 < \delta < 0.4$). The effect of each variable cannot be examined separately because they are interconnected but can be generally understood by referring to eq. (7): (i) Low $\delta$ leads to lower $\Gamma^*$, since a larger time-period is needed to reach the steady state distribution at $\tau = t^+$. (ii) High $\alpha$ (in magnitude) reduces $\partial\mathcal{H}/\partial\Gamma|_{\Gamma^*}$, due to lower injection. (iii) A combination of high $\delta$ and low $\alpha$ (in magnitude) also reduces $\partial\mathcal{H}/\partial\Gamma|_{\Gamma^*}$, since it approaches $\bar{U} = 0$.

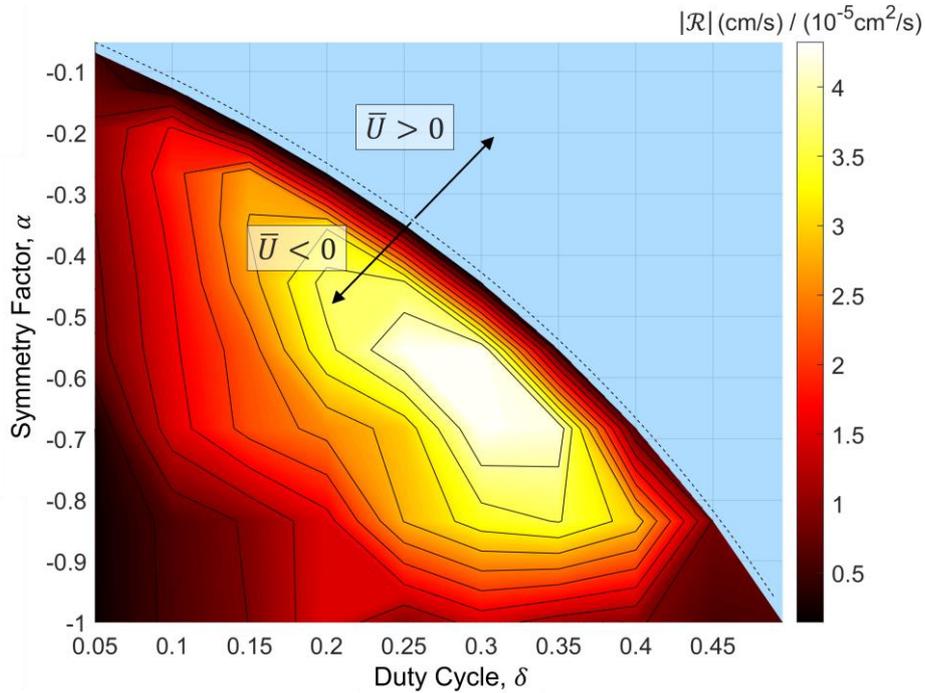

Figure 7. Separation resolution as function of the signal symmetry factor, $\alpha$, and duty cycle, $\delta$, for ratchet: $L = 1\ \mu m$, $x_c = 0.7$, $V_{max} = 2.5\ V$, and $z = 1$.

For ratchets operating near the normalized stopping frequency, the separation resolution is linear with $L^{-1}$, therefore a low period length is desired. In this work a period of $L = 1\ \mu m$ was used, although devices with significantly lower spatial periods can be fabricated. However, since the stopping frequency is proportional to $L^{-2}$, such devices will have to operate at much higher signal frequencies. The separation resolution also increases with $V_{max}$, as shown in Figure 8a. For lower voltages the increase is non-linear. This is due to the linear term of $V_{max}$ in eq. (7) and an additional increase of the slope of the characteristic velocity function, $\partial\mathcal{H}/\partial\Gamma|_{\Gamma^*}$, as shown in the inset. A resolution of $\mathcal{R} = 12\ (cm/s)/(10^{-5} cm^2/s)$ is obtained with an amplitude

of $V_{max} = 5\ V$ (and would increase for higher amplitudes). Thus, for two monovalent ions with a difference in diffusion coefficients of $\Delta D = 0.01 \cdot 10^{-5}\ cm^2/s$, which for the prevalent ions in water is a relative difference of about 1%, the maximum separation velocity is $\Delta v_{net} = \mathcal{R} \cdot \Delta D = 0.12\ cm/s$. The effect of the sawtooth sharpness, $x_c$, on the normalized net velocity is shown in Figure 8b. Moderate sharpness, such as $x_c = 0.6$, exhibit the highest separation resolution, as well as a relatively narrow frequency range, with a high ion net velocity that resembles a resonance, centered around $\Gamma = 1.3$. When the sharpness of the sawtooth is below $x_c = 0.5$, the velocity directions are flipped, and so the normalized velocity curve of $x_c = 0.3$ is a mirror image of $x_c = 0.7$. The inset shows the separation resolution as a function of $x_c$. The behavior is antisymmetric around $x_c = 0.5$. As the sawtooth becomes sharper ($x_c$ moves towards 0 or 1) the absolute value of $\mathcal{R}$ rapidly increases, and then decreases, with an optimal value at moderate sharpness values $x_c = 0.4\ \&\ 0.6$. When the sawtooth is spatially symmetric, $x_c = 0.5$, there is no ratchet effect, hence the net ion velocity is zero.

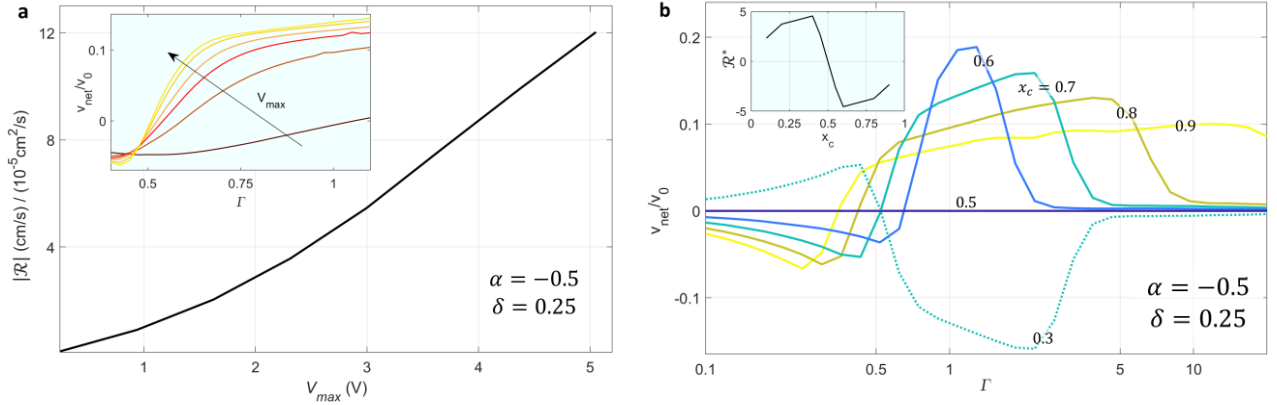

Figure 8. (a) Separation resolution (in absolute value) as a function of the maximum amplitude $V_{max}$, with $x_c = 0.7$. The inset shows the characteristic net velocity curves. (b) Characteristic net velocity curves for different $x_c$ values, with $V_{max} = 2.5\ V$. The inset shows the Separation resolution as a function of $x_c$. $\mathcal{R}^*$ is in units of $(cm/s)/(10^{-5}cm^2/s)$. Ratchet parameters for both graphs are: $L = 1\ \mu m$, $\alpha = -0.5$, $\delta = 0.25$ and $z = 1$.

Drift currents are influenced by the charge number and diffusion coefficient of the ions; therefore, when separating ions with different charge numbers, the separation ability is not determined by a difference in the diffusion coefficient alone, but by differences in their product, $zD$. Eq. (7) cannot be used to calculate the separation resolution in this case, since different ion charge numbers result in different characteristic velocity functions. Therefore, to find the separation resolution, a case-by-case method should be taken. Figure 9 shows an example of such a case for anions. Nitrate ($NO_3^-$, $D = 1.9 \cdot 10^{-5}\ cm^2/s$), which is a well-known water pollutant,[44,49] is to be extracted from water with a higher concentration of chlorine ($Cl^-$, $D = 2.03 \cdot 10^{-5}\ cm^2/s$). Carbonate ($CO_3^{2-}$, $D = 0.96 \cdot 10^{-5}\ cm^2/s$) is also shown for comparison since it has a similar $zD$ value as $NO_3^-$, and is also found in water.[43] In this case, when the ratchet is operated at 47 kHz, which is the stopping frequency of $Cl^-$, the separation velocity of the high charge number ion $CO_3^{2-}$ is much higher than of $NO_3^-$. Also note that $NO_3^-$ and $CO_3^{2-}$ can themselves be separated, even though their $zD$ values are very close. The inset of Figure 9 shows the characteristic net velocity curves for different charge numbers, and the separation resolution between same-charge number ions. The separation resolution increases significantly with the charge number, due to the linear term in eq. (7) and an increase of the slope $\partial \mathcal{H}/\partial \Gamma|_{\Gamma^*}$. Another example with common cation pollutants found in water is shown in Figure S2. Lead ($Pb^{2+}$, $D = 0.95 \cdot 10^{-5}\ cm^2/s$) can be separated from sodium ($Na^+$, $D = 1.33 \cdot 10^{-5}\ cm^2/s$) with a relatively high velocity since they have a high $zD$ difference. However, Copper ($Cu^{2+}$, $D = 0.73 \cdot 10^{-5}\ cm^2/s$) cannot be separated from sodium with high velocities, even though they have a non-negligible $zD$ difference, since their stopping frequencies coincide.

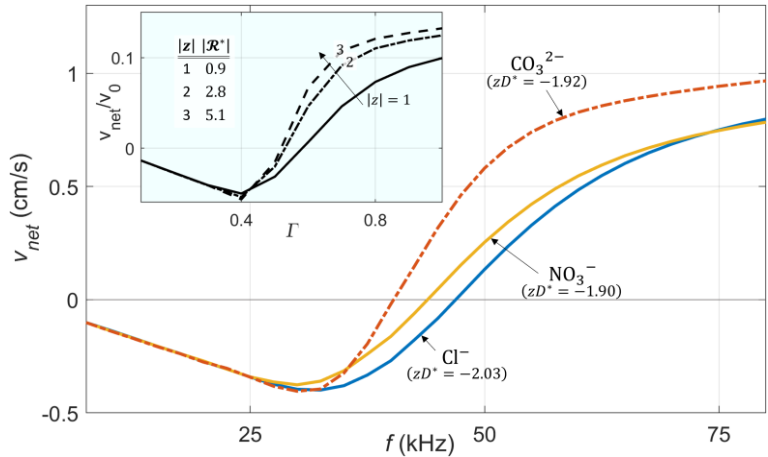

Figure 9. The net velocity of $Cl^-$, $CO_3^{2-}$, and $NO_3^-$ anions as a function of input signal frequency. The units for $zD^*$ are in $(10^{-5} cm^2/s)$. The inset shows the velocity characteristic curves and the separation resolution for different $|z|$ values. $|\mathcal{R}^*|$ is in units of $(cm/s)/(10^{-5} cm^2/s)$. Ratchet parameters: $L = 1\ \mu m$, $x_c = 0.7$, $V_{max} = -1\ V$, $\alpha = -0.5$, $\delta = 0.25$.

## Discussion

In this work we have modelled the performance of a ratchet driving ions assuming an infinitely periodic system in which no concentration gradient develops between spatial periods. This is similar to a photovoltaic cell driving current under short circuit conditions where no power is converted. However, when driving ions between two finite reservoirs, a concentration gradient develops impeding further ion transport, and the transport of ions against a force leads to a production of work. Thus, the current model can be used to describe only the ratchet operation before significant concentration gradients are formed, or a system in which the ratchet is separating two large reservoirs such that the ion concentration within them is hardly perturbed. To model a device that performs work, a concentration gradient must be considered. This can be done by assuming a per-period concentration gradient, or by solving the coupled Nernst-Poisson-Planck equations with finite reservoirs or predefined concentration gradients.

Since in the model presented the electric potential distribution is an input to the calculation, and Poisson's equation is not solved; the model does not consider the interactions between the ions, the potential being screened by them, nor the columbic forces arising when separating cations and anions. Such conditions can be obtained in extremely dilute solutions or in devices with a length scale that is smaller than the Debye length. For systems that do not comply with these assumptions, the model offers an ideal testcase that can help determine the optimal electric potential distribution for a given application. The screening of the applied potential by the solution ions reduces dramatically the amplitude of the potential modulation experienced in the solution reducing the efficiency of the device. Thus, similar to other electrically driven ion manipulation methods, such as electrodialysis and capacitive deionization, ratchet-based separation is expected to be less applicable to highly concentrated solutions. As above, an estimation of the operable concentrations and obtainable concentration gradients requires solving the coupled Nernst-Poisson-Planck equations.

The unique ability of ratchets to drive particles with the same charge in opposite directions makes them very attractive for ion separation applications. This characteristic may be most significant when targeting the removal of trace ions from a solution where conventional membrane-based separation methods require extreme levels of selectivity. For example, the total amount of lithium in the world's oceans is enough to meet the increasing global demand in an environmentally sustainable way, but its extraction is highly challenging, due to its low concentration of about 0.15 ppm.[50,51] Lithium is conveniently positioned with one of the lowest $zD$ values, so that it can be separated in a single process from most other cations that are present in ocean water at much higher concentrations, including $Mg^{2+}$, $Ca^{2+}$, $Na^+$, and $K^+$. Moreover, the interaction between the main ions and the trace ions may result in a synergistic effect that can further enhance the selectivity of a ratchet-based separation system. For example, consider a system where a ratcheting membrane is separating two closed compartments, filled with an NaCl aqueous solution with another cation in a trace amount. The ratchet pumps the $Na^+$ ions to one side and the $Cl^-$ ions to the other, until the build-up of electrostatic forces and backward diffusion negates the net force induced by the ratchet. The ratchet frequency is tuned to drive the trace ions in an opposite direction to that of the sodium ions. Since the trace ions are at a very low concentration, their contribution to the electrostatic potential is negligible. Thus, while the electrostatic potential induced by the separation of the sodium and chlorine ions impedes the transport of the main ions in the solution, it accelerates the oppositely flowing trace ions. Detailed modelling of this effect requires a full solution of the coupled Nernst-Poisson-Planck equations for all the ions in the solution and is left for future work.

For small differences in diffusion coefficients, the separation resolution increases linearly, yet as $\Delta D$ increases the separation velocity saturates. To reach higher separation velocities, the 'resonance like' behavior (obtained at moderate sawtooth sharpness such as $x_c = 0.6$) can be used, as shown in Figure 8b. For example, in some microbial electrochemical systems, such as microbial fuel cells (MFCs), operating in physiological environments, there is a great need to achieve high membrane selectivity for $H^+$ with respect to $Na^+$ and $K^+$.[52] Under the conditions described in Figure 8b with $x_c = 0.6$, the maximum velocity for $H^+$ is obtained at a normalized frequency of $\Gamma = 1.3$ which corresponds to a frequency of 1.2 MHz. For $Na^+$ and $K^+$ this frequency translates to normalized frequencies of $\Gamma = 9.1$ and $\Gamma = 6.2$, respectively. As can be seen in Figure 8b, at such high normalized frequencies the $Na^+$ and $K^+$ velocities are near zero leading to a velocity ratio of 500:1 and 350:1, relative to $Na^+$ and $K^+$, respectively.

Velocity reversal can also be used for driving opposite charge ions in the same direction, a process we term ambipolar ion transport. Figure 10 shows an example for ambipolar transport of the main ions in seawater. When the ratchet is operated at 243 kHz, both the negative and positive ions are transported in the same direction at a velocity of 2 cm/s. Since both cations and anions are driven in the same direction, cations and anions are not separated, and the electrostatic potential that opposes ion transport in other cases does not develop. This makes ambipolar ion transport an attractive mechanism for desalination and distillation applications.

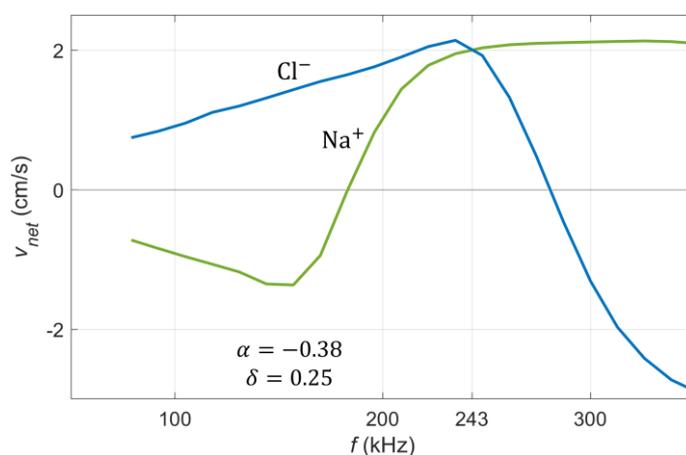

Figure 10. The net velocity of $Cl^-$ and $Na^+$ showing ambipolar transport. Ratchet parameters: $L = 1\ \mu m$, $x_c = 0.6$, $V_{max} = 5\ V$, $\alpha = -0.38$, $\delta = 0.25$.

**Conclusion**

In this work we have laid the theoretical groundwork for ratchet-driven ion separation. We have shown that a flashing ratchet can be used to drive ions with the same charge number in opposite directions based on the difference in diffusion coefficients. This unique ratchet property paves the way for rapid extraction of trace ions from solutions where conventional membrane-based separation processes do not offer sufficient selectivity. We define the ratchet separation resolution and show that under moderate voltages, prevalent ions in water with a 1% relative difference in diffusion coefficients can be driven in opposite directions with a velocity difference of 1.2 mm/s. Moreover, non-linearities in the ion velocity with the input signal amplitude and ion valence can be further utilized to enhance the ratchet selectivity. Thus, ratchet-based ion pumping systems may prove beneficial for applications such as water treatment, bio-medical devices, ion sensors, solar fuels generators and more.

**Acknowledgments**

The authors would like to acknowledge Francesca Toma for helpful discussions. AH acknowledges the support of the Boris Mints Institute. JWA was supported by the Joint Center for Artificial Photosynthesis, a DOE Energy Innovation Hub, supported through the Office of Science of the United States. Department of Energy under Award No. DE-SC0004993. SA acknowledges the support of the Gordon and Betty Moore Foundation under a Moore Inventor Fellowship (GBMF grant #5641) and The Beall Family Foundation (UCI Beall Innovation Award). GS thanks the Azrieli Foundation for financial support within the Azrieli Fellows program.

**Conflict of Interest**

GS, JWA, and SA filed patent applications US 16/907,076 and US 17/125,341 for ratchet-based ion pumping membrane systems. There are no other conflicts of interest to declare.

**Author Contributions**

Conceptualization: GS, SA

Investigation: AH, GS

Methodology: AH, JWA, GS

Data curation: AH

Formal analysis: AH

Visualization: AH, GS

Writing – original draft: AH, GS

Writing – review and editing: AH, GS, JWA, SA

Supervision, Project Administration, and Funding Acquisition: GS

## Supporting Information

### Model Validation

To validate our numerical model, we compared our simulation to the analytical solution presented by Kedem et al.[13], and developed by Rozenbaum[27] for low energy (compared to the thermal energy). The analytical solution uses a biharmonic potential distribution, given in eq. (S1):

$$V(x) = V_1 \sin(2\pi x/L) + V_2 \sin(4\pi x/L) \tag{S1}$$

Here we defined $V_1 = V_{max}$ and used $V_2 = V_{max}/5$ to get a spatially asymmetric potential, that is similar in shape to a sawtooth. The particles net velocity is calculated by eq. (S2)[13]:

$$v_{net,analytic} = \frac{\pi}{4} \frac{D}{L} \beta^3 V_1^2 V_2 (1-\alpha)^2 [(1+\alpha)\Phi_1(\Gamma,\delta) + (1-\alpha)\Phi_2(\Gamma,\delta)]$$

$$\Phi_1(\Gamma,\delta) = \frac{12}{\pi^2} xC(x,y), \quad \Phi_2(\Gamma,\delta) = \frac{12}{\pi^2} xG(x,y)$$

$$x = \beta V_{max} \Gamma/(2\pi), \quad y = 2\pi(1-\delta)$$

$$C(x,y) = \sum_{n=1}^{\infty} \frac{x(7+2x^2n^2)(1-\cos yn)}{(1+x^2n^2)^2(16+x^2n^2)}, \quad G(x,y) = \left(1 - 2\frac{y}{2\pi}\right)C(x,y) - \frac{D(x,y)}{2\pi} \tag{S2}$$

$$D(x,y) = \sum_{n=2}^{\infty} \sum_{m=1-n}^{-1} [B(x,y,n,m) + B(x,y,m,n)] + \sum_{n,m=1}^{\infty} B(x,y,n,m)$$

$$B(x,y,n,m) = \frac{x[5n+(n+m)(4-x^2n^2)][\sin yn\,(1-\cos ym) + \sin ym\,(1-\cos yn)]}{(n+m)m(1+x^2n^2)(16+x^2n^2)(1+x^2(n+m)^2)}$$

To compare our numerical model to the analytical model, simulations were performed with the potential distribution of eq. (S1). Figure S1 shows a comparison of the normalized net velocity as a function of the normalized frequency, calculated with both models, for different potential amplitudes. The two models agree well for low amplitudes, i.e., $\beta V_{max} < 1$ ($V_{max} < 25.7$ mV, @25°C), while for higher amplitudes the models diverge. This result is to be expected since the analytical model was developed with a low energy assumption. Since ion transport problems usually require higher potentials to drive a sufficient current, this emphasizes the need for a numerical solution.

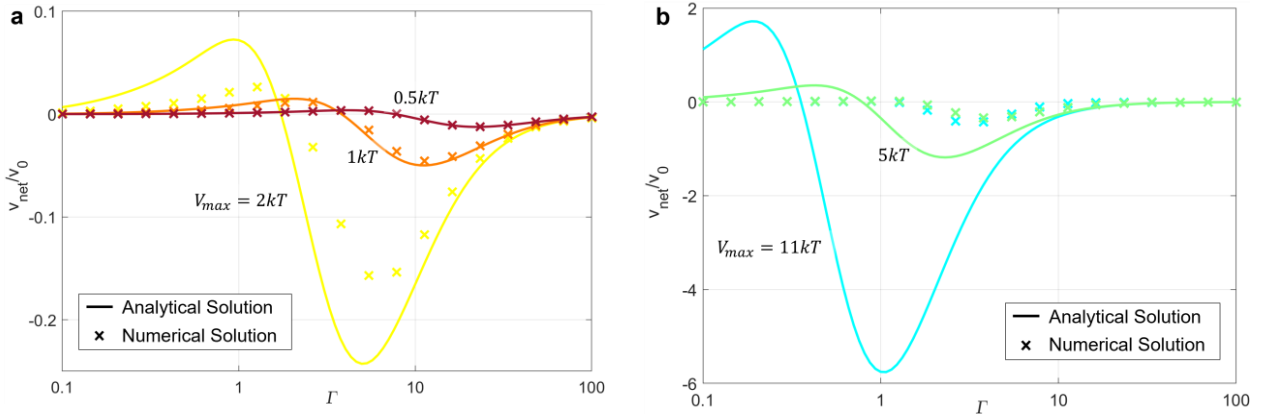

*Figure S1. Model validation. Numerical vs. analytical solution for a potential distribution described by eq. (S1), ratchet parameters: : $L = 10$ µm, $\alpha = -0.75$, $\delta = 0.25$, and (a) $V_{max}$ between $0.5kT$ to $2kT$, (b) $V_{max}$ between $5kT$ to $11kT$. For clarity $kT$ is used instead of the full expression $\beta^{-1} = k_B T_r/e$.*

### Detailed calculation of the separation resolution

We would like to find the separation resolution for any set of parameters, represented by a single characteristic net velocity curve $\mathcal{H}(\Gamma)$ and its corresponding normalized stopping frequency $\Gamma^*$. Applying the chain rule on the definition of the separation resolution, and remembering at the normalized stopping frequency, $\mathcal{H}(\Gamma^*) = 0$, we obtain:

$$\mathcal{R} = \left.\frac{\partial v_{net}}{\partial D}\right|_{\Gamma^*} = \left.\frac{\partial(v_0 \cdot \mathcal{H}(\Gamma))}{\partial D}\right|_{\Gamma^*} = \frac{\partial v_0}{\partial D}\cdot \mathcal{H}(\Gamma^*) + v_0 \left.\frac{\partial \mathcal{H}}{\partial D}\right|_{\Gamma^*} = v_0 \left.\frac{\partial \mathcal{H}}{\partial \Gamma}\right|_{\Gamma^*} \left.\frac{\partial \Gamma}{\partial D}\right|_{\Gamma^*} \tag{S3}$$

Where:

$$\frac{\partial \Gamma}{\partial D} = \frac{\partial}{\partial D}\left(\frac{fL^2}{zD\beta V_{max}}\right) = -\frac{fL^2}{\beta V_{max} zD^2} = -\frac{\Gamma}{D} \tag{S4}$$

Inserting into eq. (S3) yields:

$$\mathcal{R} = \left(\frac{zD\beta V_{max}}{L}\right)\left(-\frac{\Gamma^*}{D}\right) \cdot \left.\frac{\partial \mathcal{H}}{\partial \Gamma}\right|_{\Gamma^*} = -\frac{z\beta V_{max}}{L} \cdot \Gamma^* \left.\frac{\partial \mathcal{H}}{\partial \Gamma}\right|_{\Gamma^*} \tag{S5}$$

The separation of lead and copper ions from sodium

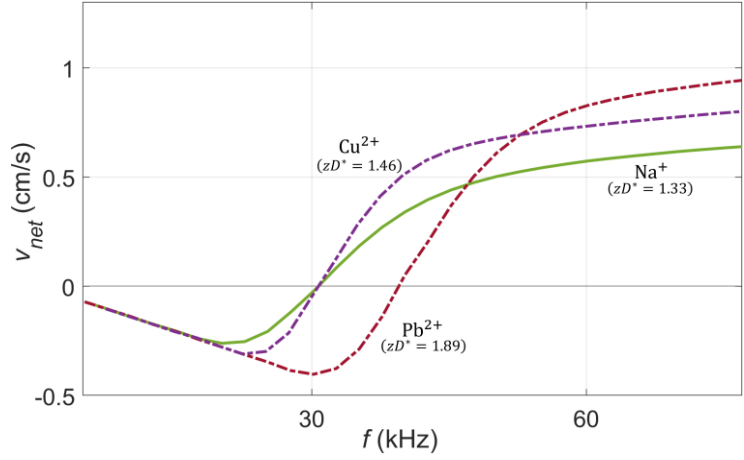

Figure S2. The net velocity of $Na^+$, $Cu^{2+}$, and $Pb^{2+}$ cations as a function of input signal frequency. The units for $zD^*$ are in $(10^{-5} cm^2/s)$. Ratchet parameters: $L = 1\ \mu m$, $x_c = 0.7$, $V_{max} = 1\ V$, $\alpha = -0.5$, $\delta = 0.25$.

**References**

1. Tang, C. & Bruening, M. L. Ion separations with membranes. *J. Polym. Sci.* **58**, 2831–2856 (2020).

2. Epsztein, R., DuChanois, R. M., Ritt, C. L., Noy, A. & Elimelech, M. Towards single-species selectivity of membranes with subnanometre pores. *Nat. Nanotechnol.* **15**, 426–436 (2020).

3. Sholl, D. S. & Lively, R. P. Seven chemical separations to change the world. *Nature* **532**, 435–437 (2016).

4. World Health Organization. *Guidelines for drinking-water quality*. (2017).

5. Tang, C., Yaroshchuk, A. & Bruening, M. L. Flow through negatively charged, nanoporous membranes separates Li+and K+due to induced electromigration. *Chem. Commun.* **56**, 10954–10957 (2020).

6. Armstrong, J. A., Bernal, E. E. L., Yaroshchuk, A. & Bruening, M. L. Separation of ions using polyelectrolyte-modified nanoporous track-etched membranes. *Langmuir* **29**, 10287–10296 (2013).

7. Yaroshchuk, A. E. Asymptotic behaviour in the pressure-driven separations of ions of different mobilities in charged porous membranes. *J. Memb. Sci.* **167**, 163–185 (2000).

8. Yaroshchuk, A. E. Dielectric exclusion of ions from membranes. *Advances in Colloid and Interface Science* **85**, 193–230 (2000).

9. Yaroshchuk, A. E. Negative rejection of ions in pressure-driven membrane processes. *Adv. Colloid Interface Sci.* **139**, 150–173 (2008).

10. Hänggi, P. & Marchesoni, F. Artificial Brownian motors: Controlling transport on the nanoscale. *Rev. Mod. Phys.* **81**, 387–442 (2009).

11. Reimann, P. Brownian motors: Noisy transport far from equilibrium. *Phys. Rep.* **361**, 57–265 (2002).

12. Lau, B. & Kedem, O. Electron ratchets: State of the field and future challenges. *J. Chem. Phys.* **152**, 200901 (2020).

13. Kedem, O., Lau, B. & Weiss, E. A. How to Drive a Flashing Electron Ratchet to Maximize Current. *Nano Lett.* **17**, 5848–5854 (2017).

14. Kedem, O., Lau, B. & Weiss, E. A. Mechanisms of Symmetry Breaking in a Multidimensional Flashing Particle Ratchet. *ACS Nano* **11**, 7148–7155 (2017).

15. Kedem, O., Lau, B., Ratner, M. A. & Weiss, E. A. Light-responsive organic flashing electron ratchet. *Proc. Natl. Acad. Sci.* **114**, 8698–8703 (2017).
16. Tanaka, T., Nakano, Y. & Kasai, S. GaAs-based nanowire devices with multiple asymmetric gates for electrical brownian ratchets. *Jpn. J. Appl. Phys.* **52**, (2013).
17. Linke, H. *et al.* Experimental tunneling ratchets. *Science (80-. ).* **286**, 2314–2317 (1999).
18. Mikhnenko, O. V., Collins, S. D. & Nguyen, T. Q. Rectifying electrical noise with an ionic-organic ratchet. *Adv. Mater.* **27**, 2007–2012 (2015).
19. Hu, Y. *et al.* Understanding the Device Physics in Polymer-Based Ionic-Organic Ratchets. *Adv. Mater.* **29**, (2017).
20. Hubmann, S. *et al.* Giant ratchet magneto-photocurrent in graphene lateral superlattices. *Phys. Rev. Res.* **2**, 33186 (2020).
21. Roeling, E. M. *et al.* Organic electronic ratchets doing work. *Nat. Mater.* **10**, 51–55 (2011).
22. Roeling, E. M. *et al.* The performance of organic electronic ratchets. *AIP Adv.* **2**, (2012).
23. Andersson, O., Maas, J., Gelinck, G. & Kemerink, M. Scalable Electronic Ratchet with Over 10% Rectification Efficiency. *Adv. Sci.* **7**, (2020).
24. Olbrich, P. *et al.* Terahertz ratchet effects in graphene with a lateral superlattice. *Phys. Rev. B* **93**, 1–15 (2016).
25. Faltermeier, P. *et al.* Helicity sensitive terahertz radiation detection by dual-grating-gate high electron mobility transistors. *J. Appl. Phys.* **118**, 084301 (2015).
26. Kabir, M., Unluer, D., Li, L., Ghosh, A. W. & Stan, M. R. Electronic Ratchet : A Non-Equilibrium , Low Power. *2011 11th IEEE Int. Conf. Nanotechnol.* 482–486 (2011). doi:10.1109/NANO.2011.6144610
27. Rozenbaum, V. M. High-temperature brownian motors: Deterministic and stochastic fluctuations of a periodic potential. *JETP Lett.* **88**, 342–346 (2008).
28. Marbach, S. & Bocquet, L. Active sieving across driven nanopores for tunable selectivity. *J. Chem. Phys.* **147**, (2017).
29. Marbach, S., Kavokine, N. & Bocquet, L. Resonant osmosis across active switchable membranes. *J. Chem. Phys.* **152**, (2020).
30. Parrondo, J. M. R., Blanco, J. M., Cao, F. J. & Brito, R. Efficiency of Brownian motors. *Europhys. Lett.* **43**, 248–254 (1998).
31. Nouri, R. & Guan, W. Nanofluidic charged-coupled devices for controlled DNA transport and separation. *Nanotechnology* **32**, 345501 (2021).
32. Ardo, S. *et al.* Ratchet-based ion pumping membrane systems. US patent application 16/907,076 (2020).
33. Rozenbaum, V. M., Korochkova, T. Y., Chernova, A. A. & Dekhtyar, M. L. Brownian motor with competing spatial and temporal asymmetry of potential energy. *Phys. Rev. E - Stat. Nonlinear, Soft Matter Phys.* **83**, 1–10 (2011).
34. Kodaimati, M. S., Kedem, O., Schatz, G. C. & Weiss, E. A. Empirical Mappings of the Frequency Response of an Electron Ratchet to the Characteristics of the Polymer Transport Layer. *J. Phys. Chem. C* **123**, 22050–22057 (2019).
35. Skaug, M. J., Schwemmer, C., Fringes, S., Rawlings, C. D. & Knoll, A. W. Nanofluidic rocking Brownian motors. *Science (80-. ).* **359**, 1505–1508 (2018).
36. Schwemmer, C., Fringes, S., Duerig, U., Ryu, Y. K. & Knoll, A. W. Experimental Observation of Current Reversal in a Rocking Brownian Motor. *Phys. Rev. Lett.* **121**, 104102 (2018).
37. Nicollier, P. *et al.* Nanometer-Scale-Resolution Multichannel Separation of Spherical Particles in a Rocking Ratchet with Increasing Barrier Heights. *Phys. Rev. Appl.* **15**, 034006 (2021).
38. Chen, Q., Ai, B. Q. & Xiong, J. W. Brownian transport of finite size particles in a periodic channel coexisting with an energetic potential. *Chaos* **24**, (2014).

39. Słapik, A., Łuczka, J., Hänggi, P. & Spiechowicz, J. Tunable Mass Separation via Negative Mobility. *Physical Review Letters* **122**, (2019).

40. Motz, T., Schmid, G., Hänggi, P., Reguera, D. & Rubí, J. M. Optimizing the performance of the entropic splitter for particle separation. *J. Chem. Phys.* **141**, 074104 (2014).

41. Reguera, D. *et al.* Entropic splitter for particle separation. *Phys. Rev. Lett.* **108**, 1–5 (2012).

42. Yang, B., Long, F. & Mei, D. C. Negative mobility and multiple current reversals induced by colored thermal fluctuation in an asymmetric periodic potential. *Eur. Phys. J. B* **85**, 2–7 (2012).

43. Luo, T., Abdu, S. & Wessling, M. Selectivity of ion exchange membranes: A review. *J. Memb. Sci.* **555**, 429–454 (2018).

44. Sata, T. Studies on anion exchange membranes having permselectivity for specific anions in electrodialysis — effect of hydrophilicity of anion exchange membranes on permselectivity of anions. **167**, 1–31 (2000).

45. Sata, T., Sata, T. & Yang, W. Studies on cation-exchange membranes having permselectivity between cations in electrodialysis. **206**, 31–60 (2002).

46. Luo, T., Roghmans, F. & Wessling, M. Ion mobility and partition determine the counter-ion selectivity of ion exchange membranes. *J. Memb. Sci.* **597**, 117645 (2020).

47. Acar, E. T., Buchsbaum, S. F., Combs, C., Fornasiero, F. & Siwy, Z. S. Biomimetic potassium-selective nanopores. 1–8 (2019).

48. Li, X. *et al.* Fast and selective fluoride ion conduction in sub-1-nanometer metal-organic framework channels. *Nat. Commun.* **10**, 1–12 (2019).

49. T. Matos, C., Velizarov, S., G. Crespo, J. o & A.M. Reis, M. Simultaneous removal of perchlorate and nitrate from drinking water using the ion exchange membrane bioreactor c ... bioreactor concept. *water Res.* **40**, (2006).

50. Yang, S., Zhang, F., Ding, H. & He, P. Lithium Metal Extraction from Seawater. *Joule* **2**, 1648–1651 (2018).

51. Swain, B. Recovery and recycling of lithium : A review. *Sep. Purif. Technol.* **172**, 388–403 (2017).

52. Harnisch, F. & Schrçder, U. Selectivity versus Mobility : Separation of Anode and Cathode in Microbial Bioelectrochemical Systems. 921–926 (2009). doi:10.1002/cssc.200900111